\documentclass[prl,floats,showpacs,aps,twocolumn,floatfix,amsmath]{revtex4}
\usepackage{graphicx}
\begin{document}
\title{Universal routes to spontaneous ${\cal PT}$-symmetry breaking\\
in non-hermitian quantum systems}

\author{Henning Schomerus}
\affiliation{Department of Physics, Lancaster University, Lancaster,  LA1
4YB, United Kingdom} \pacs{05.45.Mt, 03.65.-w, 42.25.Dd}

\date{November 2010}

\begin{abstract}
${\cal PT}$-symmetric systems can have a real spectrum even when their
Hamiltonian is non-hermitian, but develop a complex spectrum when the degree
of non-hermiticity increases. Here we utilize random-matrix theory to show
that this spontaneous ${\cal PT}$-symmetry breaking can occur via two
distinct mechanisms, whose predominance is associated to different
universality classes. Present optical experiments fall into the orthogonal
class, where symmetry-induced level crossings render the characteristic
absorption rate independent of the coupling strength between the
symmetry-related parts of the system.
\end{abstract}

\maketitle

\begin{figure}[t]
\includegraphics[width=.8\columnwidth]{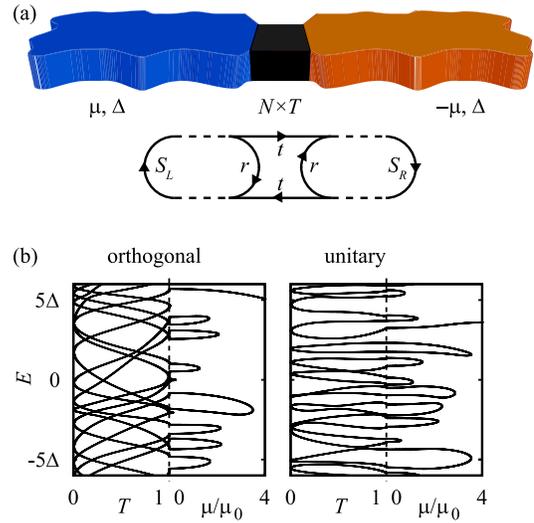}
\caption{\label{fig:1} (Color online) (a) Sketch of a nonhermitian ${\cal PT}$-symmetric system, where a region with absorption
rate $\mu$ (and mean level spacing $\Delta$, left) is coupled symmetrically via
a tunnel barrier (supporting $N$ channels with transmission probability $T$) to an amplifying region with a matching amplification rate (right).
Below this, the scattering description of the system. (b) Two routes to spontaneous ${\cal PT}$-symmetry breaking,
depending on whether the hermitian limit  $\mu=0$ is ${\cal T}$-symmetric
(orthogonal class displaying level crossings, left) or not (unitary class displaying avoided crossings, right).
Shown are  real eigenvalues of a random Hamiltonian ${\cal H}$ [Eq.\ (\ref{eq:h})] as function of $T$ for fixed $\mu=0$ (left of dashed line), and then as a function of $\mu$
for fixed $T=1$ (right of dashed line). Complex-valued levels (formed by level coalescence at $\mu>0$) are not shown. Here $\mu_0=\sqrt{N} \Delta/2\pi$, and we set $N=10$.
}
\end{figure}

Non-hermitian quantum systems generally have a complex energy spectrum, with
imaginary parts of the energies related to decay or amplification rates.
However, when loss and gain are in balance the spectrum can still be real.
One intensely researched route to try and achieve such a balance is to couple
two identical systems symmetrically and then induce opposite amounts of gain
and loss into the two parts, as illustrated in Fig.\ \ref{fig:1}(a)
\cite{bender,znojil,benderbrody,mostafazadeh,benderreview,ptscattering,makris,longhi,experiments,hs,laser-absorber,kottos,graefe}.
The Hamiltonian then possesses a combined parity (${\cal P}$) and
time-reversal (${\cal T}$) symmetry, and its secular equation is real.
However, this does not guarantee a real spectrum; as the level of
non-hermiticity (loss and gain) is increased, pairs of complex-conjugate
energy levels appear \cite{benderbrody,mostafazadeh,benderreview}. This
phenomenon of spontaneous ${\cal PT}$-symmetry breaking has gained recent
prominence because it leads to optical effects such as double refraction,
solitons and non-reciprocal diffraction patterns,  which provide mechanisms
for the design of unidirectional couplers and left-right sensors
\cite{makris,longhi},  concepts that are now being realized experimentally in
a variety of optical settings \cite{experiments}. Over the past months, these
systems were proposed for at-threshold lasers \cite{hs} and laser-absorbers
\cite{laser-absorber,absorber}. In turn, these
developments have instigated a deeper theoretical understanding of the role
of the dynamics (such as the consequences of Anderson localization and wave chaos \cite{kottos},
as well as interactions \cite{graefe}). In this paper, we establish
distinct universality classes which directly affect the nature of spontaneous
${\cal PT}$-symmetry breaking.

\begin{figure}[t]
\includegraphics[width=\columnwidth]{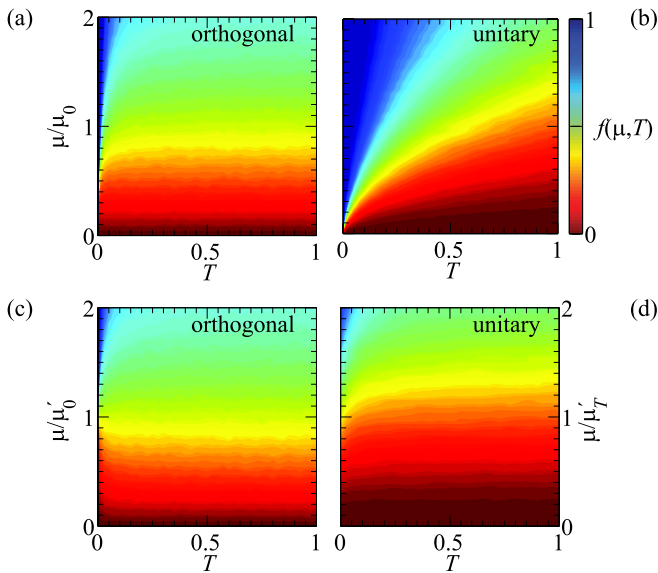}
\caption{\label{fig:2} (Color online)
Gradient plots of the ensemble-averaged fraction $f(\mu,T)$ of complex energy levels (among all levels within a range of
energies over which the mean  spacing $\Delta$ can be assumed constant), for the orthogonal class (left) and the unitary class (right).
In (a) and (b), $\mu$ is scaled to $\mu_0$.
The bottom panels show same data with $\mu$ scaled to $\mu_0'=\mu_0/\sqrt{1+1/NT}$ (c) and $\mu_T'=\sqrt{T}\mu_0$
(d). Numerical results with $N=50$.
}
\end{figure}

To do so, we derive random-matrix ensembles where loss and gain in the two
parts of the system are implemented by a uniform rate $\mu$, while coupling
is established through $N$ channels with transmission probability $T$; the
mean level spacing of the decoupled parts is $\Delta$ \cite{jaincomment}. We
find that the mechanism behind spontaneous ${\cal PT}$-symmetry breaking
depends on whether the hermitian limit $\mu=0$ is time-reversal symmetric or
not, amounting to a predominance of symmetry-induced level crossings or level
repulsion, as illustrated in Fig. \ref{fig:1}(b). This results in different
characteristic scales $\mu_{PT}$ of amplification/absorption governing the
transition from an essentially real to an essentially complex spectrum.
Present optical experiments and theoretical studies concern systems without
magneto-optical effect (the orthogonal symmetry class), in which the
hermitian limit is time-reversal symmetric. In this case $\mu_{PT}\sim
\sqrt{N} \Delta/2\pi\equiv \mu_0$ becomes \emph{fully independent of the
coupling strength} as soon as $T$ surpasses a parametrically small threshold
$T_c \sim 1/N$, thereby exhibiting a level of universality that goes beyond
what is normally encountered in mesoscopic systems. For weak coupling
($T<T_c$), $\mu_{PT}\sim \sqrt{NT} \mu_0$. Adding magneto-optical effects to
the system essentially changes the nature of the transition. In this case
(the unitary symmetry class), $\mu_{PT}\sim \sqrt{T}\mu_0$ in the full range
of weak and strong coupling. These findings are illustrated in Fig.\
\ref{fig:2}.

\emph{Random-matrix ensembles}.---To derive the appropriate random-matrix
ensembles we formulate a quantization condition based on scattering theory
\cite{hs,beenakkerreview,frahm}. The $N\times N$-dimensional scattering matrix
\begin{equation}
S_L(E;\mu) = 1-2i V^\dagger (E-i\mu-H+i VV^\dagger)^{-1}V
\end{equation}
of the left subsystem can be expressed in terms of an $M\times M$-dimensional
Hamiltonian $H$, which is real symmetric (a member of the standard Gaussian
orthogonal ensemble, GOE) if the hermitian limit $\mu=0$  is time-reversal
symmetric, and complex hermitian (a member of the  Gaussian unitary ensemble,
GUE) if this is not the case \cite{mehta}. We assume $M\gg N
\gg 1$ and denote the mean level spacing in the energy range of interest by
$\Delta$. The $M\times N$ coupling matrix $V$ then fulfills $VV^\dagger={\rm
diag}{(v_m)}$, where $N$ diagonal entries $v_m=\Delta M/\pi$ correspond to
fully transparent channels, and $M-N$ entries $v_m=0$ describe the closed
channels \cite{beenakkerreview}. Adopting a basis where the time-reversal operation ${\cal T}$ is
identical to complex conjugation,
${\cal PT}$-symmetry results in the relation \cite{hs}
\begin{eqnarray}
S_R(E;-\mu)&=&[S_L^{-1}(E^*;\mu)]^*\nonumber \\ &=& 1-2i V^\dagger (E+i\mu-H^*+i VV^\dagger)^{-1}V\quad
\end{eqnarray}
for the scattering matrix of the right subsystem. The tunnel barrier is
described by reflection amplitudes $r=-\sqrt{1-T}$ and transmission
amplitudes $t=i\sqrt{T}$.
As shown in the lower part of Fig.\ \ref{fig:1}(a), these scattering matrices relate amplitudes of left and right
propagating waves at the two interfaces of the tunnel barrier. The
requirement of consistency of these relations results in the quantization
condition
\begin{equation}
{\rm det}\,\left[\left(\begin{array}{cc}r & t \\t & r \\ \end{array}\right)\left(\begin{array}{cc}S_L & 0 \\0 & S_R \\ \end{array}\right)-\openone\right]=0,
\end{equation}
which can be rearranged into an eigenvalue problem ${\rm det}\,(E-{\cal H})=0$ with  effective Hamiltonian
\begin{equation}
\label{eq:h}
{\cal H}=\left(\begin{array}{cc}H-i\mu &   \Gamma  \\   \Gamma & H^*+i\mu \\ \end{array}\right).
\end{equation}
The positive semi-definite coupling matrix $\Gamma={\rm diag}\,(\gamma_m)$
now incorporates the finite transmission probability of the barrier; its $N$
non-vanishing entries read $\gamma_m=[\sqrt{T}/(1+\sqrt{1-T})]\Delta
M/\pi\equiv \gamma$ \cite{gammacomment}.

\emph{Numerical evaluation}.--- Before engaging in an analytical discussion
of the different routes to spontaneous ${\cal PT}$-symmetry breaking
[illustrated in Fig.\ \ref{fig:1}(b)], we put forward numerical results which
illustrate the physical consequences of the points to be made below. These
results, presented as (color) gradient plots in Fig.\ \ref{fig:2}, concern
the fraction $f(\mu,T)$ of complex-valued energy levels within a range where
the mean level spacing can be assumed constant. We fix $M=1000$, which ensures a large number of levels within the range in question,
and set $N=50$ \cite{numericsnote}.

Figure \ref{fig:2}(a) shows results for the orthogonal ensemble, with $\mu$
scaled to $\mu_0=\sqrt{N}\Delta/2\pi$. We see that above a small threshold
$T_c$ (to be determined below  as $T_c\sim 1/N$), the transition from the
real spectrum ($f=0$, obtained for $\mu=0$) to a spectrum which is partially
real and partially complex  spectrum ($f\sim 1/2$) indeed occurs on the scale
$\mu_{PT}\sim\mu_0$, and then is independent of the value of $T$. Only for
$T<T_c$, $\mu_{PT}\sim\sqrt{NT}\mu_0\equiv \mu_T$ is
coupling-dependent. In order to get a unified view over both regimes, we plot
in panel (c) the same data, but with $\mu$ scaled to
$\mu_0'=\mu_0/\sqrt{1+1/NT}$, which interpolates between $\mu_T$ for $T\ll
T_c$ and $\mu_0$ for $T\gg T_c$. The convergence of gradient lines for $T\to
0$ indicates that for weak coupling the transition becomes more abrupt.

Figure \ref{fig:2}(b) shows the corresponding results for the unitary
ensemble, where $\mu$ is again scaled to $\mu_0$. Here we find that a
systematic $T$-dependence persists across the full range of coupling
strengths. As shown in panel (d), this dependence takes the form
$\mu_{PT}\sim\sqrt{T}\mu_0\equiv\mu_T'$. Now, the only difference between
the strong and weak coupling regimes is a factor of order 1. In further
contrast to the orthogonal case, for weak coupling the transition remains
smooth; however, since $\mu_T'=\mu_T/\sqrt{N}\ll \mu_T$, it then occurs at a
far smaller deviation from hermiticity.

\emph{Underlying mechanisms}. We now show that the features reported above
originate from two distinct mechanisms of spontaneous ${\cal PT}$-symmetry
breaking.

It is instructive to start in a regime which can be treated perturbatively.
For $\mu=0$,  the effective Hamiltonian ${\cal H}$ [Eq.\ (\ref{eq:h})] is
hermitian and all its eigenvalues are real.  For $T=0$ ($\Gamma=0$), on the
other hand, the spectrum is a superposition of two level sequences
$E_k=\varepsilon_k\pm i\mu$, which are all complex if $\mu\neq 0$; here
$\varepsilon_k$ are the eigenvalues of $H$. Therefore, in regard to the
question of how many levels are complex, the limits $T$, $\mu\to 0$ do not
commute. Nonetheless, for $T=\mu=0$ the spectrum reduces to the superposition
of two degenerate level sequences $\varepsilon_k$, so that quasi-degenerate
perturbation theory applies. Denote by $\psi_m^{(k)}$ the wave function of
$H$ corresponding to eigenvalue $\varepsilon_k$; in random-matrix theory,
this is a random normalized vector with average $\overline{|\psi_m^{(k)}|^2
}=1/M$. Reduced to the symmetric and antisymmetric extension of this
wavefunction across the whole system, the effective Hamiltonian takes the
form
\begin{equation}
{\cal H}'=\left(\begin{array}{cc}\varepsilon_k-i\mu &   \sum_m[\psi_m^{(k)}]^2\gamma_m  \\   \sum_m[\psi_m^{(k)}]^{*2}\gamma_m  &\varepsilon_k+i\mu \\ \end{array}\right),
\end{equation}
whose eigenvalues become complex for
$\mu_{PT}=|\sum_m[\psi_m^{(k)}]^2\gamma_m|$. Therefore, on average (and using
$\gamma\sim \sqrt{T}\Delta M/2\pi$ for $T\ll 1$)
\begin{equation}
\label{eq:result1} \mu_{PT}\sim\left\{
\begin{array}{ll}
N \sqrt{T}\Delta/2\pi =\mu_T & \hbox{(orthogonal class),} \\
\sqrt{NT}\Delta  /2\pi=\mu_T' & \hbox{(unitary class),}
\end{array}
\right.
\end{equation}
which recovers the numerical scales in the weak coupling regime.

Note that the two expressions for $\mu_{PT}$ differ by the parametrically
large factor  $\sim\sqrt{N}$. Mathematically, this arises because
$\psi_m^{(k)}$ is real in the orthogonal class and complex in the unitary
class; physically, it amounts to vastly different tunnel splittings. This
difference signifies that in the orthogonal class, the levels of the
originally degenerate sequence $\varepsilon_k$ from the two subsystems
quickly cross as $T$ is increased. A second route to ${\cal PT}$-symmetry
breaking then becomes available, which involves two energy levels that are
non-degenerate for $T=0$. In order to describe this case we reformulate the
problem by starting with $\mu=0$, and exploit the thus-emerging  ${\cal
P}$-symmetry in the orthogonal class to transform the effective Hamiltonian
to
\begin{equation}
{\cal H}_{\cal P}=\left(\begin{array}{cc}H+\Gamma &   i\mu  \\   i\mu & H-\Gamma \\ \end{array}\right).
\label{eq:hp}
\end{equation}
We denote by $\varepsilon_k^{\pm}$ the two level sequences of $H\pm\Gamma$.
Since $\Gamma$ is positive semidefinite these sequences arise from the
sequence $\varepsilon_k$ by an oppositive shift which is approximately rigid.
From the resulting combined sequence, consider two levels $\varepsilon_k^+$
and $\varepsilon_l^-$ which lie adjacent to each other; the corresponding
eigenvectors are $\psi^{(k+)}$ and $\psi^{(l-)}$. Finite $\mu$ mixes these
levels, which is embodied in the reduced Hamiltonian
\begin{equation}
{\cal H}''=\left(\begin{array}{cc}\varepsilon_k^+ &   i\mu\langle \psi^{(k+)}|\psi^{(l-)}\rangle  \\  i\mu\langle \psi^{(l-)}|\psi^{(k+)}\rangle   &\varepsilon_l^- \\ \end{array}\right).
\end{equation}
Now, treating $2\Gamma$ as a perturbation which connects the $+$ and $-$
sequence, $\langle \psi^{(k+)}|\psi^{(l-)}\rangle\approx
\frac{\langle\psi^{(k+)} |2\Gamma|\psi^{(l+)}
\rangle}{\varepsilon_k^+-\varepsilon_l^+}$. Because
$|\varepsilon_k^+-\varepsilon_l^-|={\cal O}(\Delta)$ is small compared to the
shift due to the coupling, the denominator can be estimated as
$\varepsilon_k^+-\varepsilon_l^+\approx\varepsilon_l^--\varepsilon_l^+
\approx -{\langle\psi^{(l+)} |2\Gamma|\psi^{(l+)} \rangle}$. The coupling
strength drops out, and on average $|\langle
\psi^{(l-)}|\psi^{(k+)}\rangle|^2\sim 1/N$, i.e., the mixing is small. As a
result, the level pair in question becomes complex for $\mu^2 |\langle
\psi^{(l-)}|\psi^{(k+)}\rangle|^2\sim (\varepsilon_k^+-\varepsilon_l^-)^2\sim
\Delta^2$, i.e.,
\begin{equation}
\label{eq:result2} \mu_{PT} \sim \sqrt{N}\Delta  /2\pi=\mu_0  \quad(\hbox{orthogonal class, $T\gtrsim 1/N$}).
\end{equation}
As indicated, comparison of this expression with Eq.\ (\ref{eq:result1})
implies that this mechanism becomes favorable around $T= T_c\sim 1/N$.

This analysis of strong coupling  does not apply to the unitary case, which
does not display ${\cal P}$-symmetry for $\mu=0$. In the ${\cal P}$-basis, in
place of Eq.\ (\ref{eq:hp}) we then have
\begin{equation}\label{eq:hp2}
{\cal H}_{\cal P}=\left(\begin{array}{cc}{\rm Re} H+\Gamma &  i{\rm Im} H +i\mu  \\  i{\rm Im} H+i\mu & {\rm Re} H-\Gamma \\ \end{array}\right).
\end{equation}
Consequently, finite coupling not only results in a far reduced systematic
shift of the levels, but also  in a direct mixing of levels in the individual
sequences. Therefore, instead of level crossings one encounters level
repulsion. This difference is illustrated in Fig.\ \ref{fig:1}(b), which shows the
evolution of energy levels as $T$ is increased from $0$ to $1$ (while
$\mu=0$), and the subsequent fate of real levels as $\mu$ is
increased from 0 to $4\mu_0$ (while $T=1$); pairwise coalescing levels become complex, and then are no longer shown. In the orthogonal class,
such pairs trace back to well-separated levels $\varepsilon_k^+$,
$\varepsilon_l^-$ from the two different sequences (which are distinguished
by the opposite slopes of the levels for increasing coupling).
In contrast, in the unitary class the coalescing levels trace back to
originally closely spaced or degenerate levels, even when the coupling is
strong.

Finally, we point to an alternative scenario where the coupling-independent $\mu_{PT}\sim\mu_0$ becomes relevant
even in the unitary symmetry class. Observe that by definition, for a
hermitian Hamiltonian the ${\cal T}$ operation (complex conjugation) is
equivalent to transposition, an operation that we denote by ${\cal T}'$.
However, for non-hermitian systems there is a physical difference: Absorption
and amplification, taken by themselves, break ${\cal T}$-symmetry, but
preserve ${\cal T}'$-symmetry; the latter is broken by magneto-optical
effects. We find that a combined $P{\cal T}{\cal T}'$-symmetry still results
in a spectrum with levels that are either real or occur in complex conjugate
pairs. Compared to the case of ${\cal PT}$-symmetry, experimental
implementation simply requires to invert the magneto-optical effects in one
part of the system. The effective Hamiltonian then takes the form
\begin{equation} {\cal H}=\left(\begin{array}{cc}H-i\mu & \Gamma  \\   \Gamma
& H+i\mu \\ \end{array}\right),
\end{equation}
which differs from (\ref{eq:h}) when $H$ is complex (i.e., in the unitary
symmetry class). In the parity basis, the Hamiltonian takes the form of Eq.\
(\ref{eq:hp}) even for unitary symmetry. Coupling now induces level
crossings, and the transition to the complex spectrum is governed by the same
characteristic scales $\mu_T$ and $\mu_0$  as encountered in the orthogonal
symmetry class of $P{\cal T}$-symmetric systems.

\emph{Conclusions}.---In summary, we identified two routes to the formation
of complex energy levels in non-hermitian quantum systems with ${\cal
PT}$-symmetry  (spontaneous ${\cal PT}$-symmetry breaking). The predominant
mechanism depends on whether or not the hermitian limit possesses
time-reversal symmetry (orthogonal or unitary universality class,
respectively). Present optical experiments fall into the orthogonal class,
where level crossings result in a  characteristic absorption/amplification
rate $\mu_{\cal PT}$ which is independent of the coupling between the
symmetry-related parts of the system (unless the coupling is very weak). The
unitary class features strong level repulsion, which reduces $\mu_{\cal PT}$
and makes it coupling-dependent. While we employed random-matrix theory,
these findings can be verified for individual systems by varying the coupling
between their symmetry-related parts.


\begin{thebibliography}{99}
\bibitem{gammacomment}  In Ref.\ \cite{frahm}, which
    concerns proximity-induced mesoscopic superconductivity
    (where particle-hole symmetry replaces ${\cal PT}$-symmetry), the expression for $\gamma$ differs
    because there a barrier of transmission probability $T_N=2\sqrt{T}/(1+\sqrt{T})$ is traversed twice, first by an electron and
    then by the Andreev-reflected hole.
    This is only a matter of definition
    since the relevant quantities (Thouless energy $E_T\equiv \mu_T$ and dimensionless conductance $G=NT=N[T_N/(2-T_N)]^2$)
    change accordingly \cite{beenakkerreview}.
\bibitem{numericsnote} We certified that results are robust when $M$ is
    further increased. As long as $1\ll N \ll M$, they only depend on the
    number of channels $N$ inasmuch as this affects the border $T_c$
   between strong and weak coupling.
\end{thebibliography}
\end{document}